# NONPERTURBATIVE CALCULATION OF GREEN AND VERTEX FUNCTIONS IN TERMS OF PARTICLE CONTOURS

N. G. STEFANIS

*Institut für Theoretische Physik II, Ruhr-Universität Bochum*
*D-44780 Bochum, Germany*

ABSTRACT

The infrared regime of fermionic Green and vertex functions is studied analytically within a geometric approach which simulates soft interactions by an *effective* theory of contours. Expanding the particle path integral in terms of dominant contours at large distances, all-order results in the coupling constant are obtained for the renormalized fermion propagator and a universal vertex function with physical characteristics close to those associated with the Isgur-Wise function in the weak decays of heavy mesons. The extension to the ultraviolet regime is scetched.

## 1. Conceptual Skeleton

This article presents some applications of a geometric formalism, relying on (Euclidean) particle path integrals, which emulates soft interactions in fermionic Green and vertex functions by means of dominant contours in the relevant regimes of dynamics [1,2]. Expanding in terms of contours enables the evaluation of all-order expressions in the coupling constant, hence transcending perturbation theory. As a result, typical problems encountered in resumming Feynman graphs are avoided.

The method discussed below represents, in some aspects, an alternative concept to recent superparticle approaches (see, e.g., [3]) and to calculations within the heavy quark effective theory (see, e.g., [4]). All results are derived in the framework of an *effective* Abelian gauge theory (QED), in which the infraparticle [5] (i.e., fermion) mass $m$ serves as a dividing line between the hard ("heavy") and the soft ("light") degrees of freedom (see Table 1). This enables one to trade infrared (IR) divergences in the full theory for ultraviolet (UV) divergences in the low-energy effective theory.

Table 1. Connection between geometrical approach and conventional quantum field theory

| Geometry (Contours) | Physics (Feynman graphs) |
|---|---|
| Open smooth contour | Infrapropagator (self-energy and vertex corrections) |
| Fractal contour | Fermion Green function in UV regime (conjecture) |
| Self-intersecting contour with infinitesimal loop | Infraparticle vertex function with soft boson "cloud" transported intact through the interaction point (*exclusive* form factor) |
| Contour with cusp where four-velocity jumps | Infraparticle vertex giving rise to bremsstrahlung (*semi-inclusive* form factor) |
| Contour end-points | Multiplicatively renormalizable singularities |
| Contour cusps | Angle-dependent anomalous dimensions |

## 2. Contour Representation of Fermionic Systems

The main ingredients of the present approach can be summarized as follows [6,7]:

- Recast a quantum field-theoretical system into particle-based language, i.e., convert path integrals over fields into those over particle contours (using a Euclidean metric) within a Feynman-Schwinger framework.

- Incorporate a spacetime *built-in* resolution scale $\alpha$ in the initial field theoretical casting. Discretized copies of Euclidean manifolds $I\!R^d$ along contours in $I\!R \otimes I\!R^d$ are related by an averaging operator with a rapidly decreasing kernel within the discretization range (i.e., nonlocality is confined within the volume $\alpha^d$). The point to notice is that discretization emerges as *averaging* over cells (in analogy to Kadanoff's block-spin renormalization [8]) rather than as "latticization" of the space-time continuum. Hence, typical problems of lattice formulations, like fermion dubbling, etc., are not encountered, and the continuum limit can be taken at every stage of the calculation.

- Use the *geometry* of contours as guiding principle in emulating quantum field interactions by appropriate contour configurations (see, Fig. 1).

Following [7], the action of the averaging operator is defined by

$$\left(\slashed{D}^D + m^D\right)\psi(n\alpha) = -\int d^d y \left[(\slashed{\partial} - m)f(|y|)\right] U\left(L_{n\alpha,n\alpha+y}\right)\psi(n\alpha + y), \quad (1)$$

where $m$ is the bare fermion mass and $f(|y|)$ is the averaging kernel with $f(|y|) \xrightarrow{\alpha \to 0} \delta(y)$, which is, for instance, satisfied by a Gaussian distribution. Gauge invariance is manifest due to the non-integrable phase factor

$$U(L_{x,x+y}) \equiv \mathcal{P}\exp\left[-ig\int_{L_{x,x+y}} A_\mu(z)\,dz_\mu\right]. \quad (2)$$

Closed-form expressions for the generating functional or Green and vertex functions involve matrix elements of the evolution operator $U(T) = e^{-iHT} = e^{-(\slashed{D}+m)T}$ expressed in terms of particle eigenstates with respect to Fock's "fifth parameter" $T$ (e.g., Schwinger's proper time). Then the quantum mechanical Green function for spinor propagation in discretized form is given by

$$G^D(n\alpha, m\alpha|A) = \left\langle n\alpha \left|\left(\slashed{D}^D + m^D\right)^{-1}\right| m\alpha\right\rangle = \lim_{c\to 0}\int_c^\infty dT\,\left\langle n\alpha\left|e^{-iHT}\right|m\alpha\right\rangle \quad (3)$$

which can be converted into a path-integral expression to read

$$G(x,y|A) = \lim_{c\to 0}\int_c^\infty dT \int_{\substack{x(0)=x \\ x(T)=y}} [dx(\tau)][dp(\tau)]\exp\left[i\int_0^T d\tau\, p(\tau)\cdot\dot{x}(\tau)\right]$$
$$\times \mathcal{P}\exp\left\{-\int_0^T d\tau\,[i\gamma\cdot p(\tau) + m]\right\}\exp\left\{ig\int_0^T d\tau\,\dot{x}(\tau)\cdot A\,[x(\tau)]\right\}. \quad (4)$$

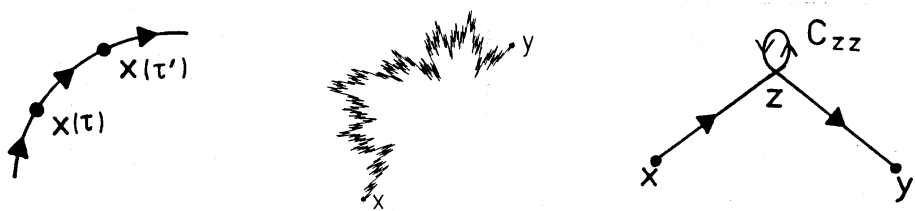

Figure 1: Contours relevant for different regimes of dynamics. The left picture shows a smooth contour which accounts for the fermion Green function in the IR regime. The fractal curve in the center illustrates quantum mechanical particle motion in the background of violent field fluctuations (UV regime). The right picture shows a self-intersecting contour with an infinitesimal loop that gives rise to a universal infraparticle vertex function.

As a result, the full fermion Green function in terms of particle path integrals reads

$$G(x,y) = \int_c^\infty dT \int_{\substack{x(0)=x \\ x(T)=y}} [dx(\tau)][dp(\tau)] \exp\left[i\int_0^T d\tau\, p(\tau)\cdot \dot{x}(\tau)\right]$$
$$\times \mathcal{P}\exp\left\{-\int_0^T d\tau[i\gamma\cdot p(\tau) + m]\right\}\left\langle \exp\left\{ig\int_0^T d\tau\, \dot{x}(\tau)\cdot A\left[x(\tau)\right]\right\}\right\rangle_A, \quad (5)$$

where the expectation value of the boson (Wilson) line exponential is

$$\left\langle \exp\left\{ig\int_0^T d\tau \dot{x}(\tau)\cdot A\left[x(\tau)\right]\right\}\right\rangle_A = \int [dA_\mu(x)]\det\left[G^{-1}(x,y|A)\right]$$
$$\times \exp\left\{ig\int_0^T d\tau\, \dot{x}(\tau)\cdot A\left[x(\tau)\right]\right\}\exp\left\{-\frac{1}{2}\int d^d z A_\mu(z)\mathcal{D}_{\mu\nu}^{-1}(\lambda)A_\nu(z)\right\}. \quad (6)$$

Eq. (5) is the particle-based version of the one-fermion Green function in field theory:

$$G(x,y) = \int [dA_\mu]\exp\left\{-\frac{1}{2}\int d^d z\, A_\mu(z)\mathcal{D}_{\mu\nu}^{-1}(\lambda)A_\nu(z)\right\}\det\left[G_D^{-1}(x,y|A)\right]G^D(x,y|A), \quad (7)$$

where the Grassmann integrations over fermion field variables have been carried out.

### 3. IR Structure of the Fermion Propagator

As we will be interested in the evaluation of Eq. (5) in the IR regime, it is sufficient to restrict attention to uniformly curved smooth contours subject to the condition that points close in configuration space are also close in parameter space, i.e., $\tau \approx \tau'$ if $x(\tau) \approx x(\tau')$ (see Fig. 1). Physically, this means that all bosonic modes with wavelengths larger than $m^{-1}$ (large scale of the system) are absorbed into the boson-fermion composite effective field and are invisible. Correspondingly, in the limit $(k/m) \to 0$ ($k$ being the boson virtuality, i.e., the small scale of the system), changes in the fermion four-velocity $\propto k/m$ vanish and fluctuations of the fermion momentum around the mass shell are of order $k \to 0$, so that the (composite) fermion acquires a "heaviness" and the particle-based description becomes justified.

The extension of this approach to the UV regime $((k/m) \to \infty)$ is seemingly at odds with the common assumption that if a contour is very complex, it can hardly be associated with a particle's trajectory – a semiclassical notion. However, one may associate the motion of a charged fermion in the background of violent field fluctuations to a *fractal* contour (see Fig. 1) that is nowhere differentiable ($|\dot{x}_\mu| \to \infty$). Such contours yield additional divergences the renormalization of which introduces angle-dependent anomalous dimensions [9,10]. In addition, pair-creation/annihilation effects must be accommodated in the formalism. In our approach, local contour discontinuities caused by such effects can be viewed as taking place at time scales smaller than the evolution pace $\epsilon$ that links successive discretized copies of $\mathbb{R}^d$. Then, by appropriately gauging the spacetime resolution scales $\alpha$ and $\epsilon$ ($\alpha/\epsilon \to 1$ for $\alpha, \epsilon \to 0$), and by readjusting the coupling constant (and mass), discontinuous contours are out of harm's way at distances larger than the discretization constants. This smoothing procedure is somewhat formal because the burden is now placed on properly defining the new effective action emerging from this renormalization-group (RG) type transformation. It is understood that vacuum polarization effects will be accounted for in the fermion determinant.

Implementing the smoothness criterion, the boson line exponential yields $\exp\{-g^2 \int_0^T d\tau \int_\tau^T d\tau' \dot{x}_\mu(\tau) \dot{x}_\nu(\tau') \langle 0|A_\mu[x(\tau)]A_\nu[x(\tau')]|0\rangle\}$. The boson correlator contains short-distance singularities $\propto |x(\tau') - x(\tau)|^{-n}$ with $n = 2, 4$ when the two contours $x(\tau)$ and $x(\tau')$ approach each other. To render the integrals finite, the "ribbon regularization" technique [11] is employed, which replaces contours by the edges of an untwisted and unwrinkled stretched-out ribbon, the width $b$ of which plays the role of an UV-regulator: $x_\mu(\tau) \to x_\mu(\tau) + b n_\mu(\tau)$, $n_\mu(\tau)$ being a unit vector normal to the contour direction and lying on the ribbon's surface. Note that the divergent parts of the correlator do not depend on derivative terms, so that they can be absorbed into an overall multiplicative renormalization constant. Taylor-expanding contours, yields $x_\mu(\tau') \simeq x_\mu(\tau) + (\tau' - \tau) \dot{x}_\mu(\tau)$ (with $|\dot{x}| = 1$), and the result for the boson line exponential reads

$$\left\langle \exp\left\{ig \int_0^T d\tau\, \dot{x}_\mu(\tau) A_\mu\left[x(\tau)\right]\right\}\right\rangle_A^{reg} = \exp\left(-\alpha \frac{T}{2b} + a \ln \frac{T}{b} + \text{finite terms}\right) \quad (8)$$

from which we infer how the physical mass $m_{ren}$ varies with the change in couplings and cutoffs: $m_r \stackrel{b \to 0}{=} m(b) + \frac{\alpha}{2b}$. The logarithmic term entering Eq. (8), furnishes a wave function renormalization constant $Z_2$, which gives rise to an anomalous dimension $\gamma_F$ for the fermion field. Introducing a renormalization mass scale $\mu$ by $\ln \frac{T}{b} = \ln(T\mu) - \ln(b\mu)$, we find the following all-order expressions (in the fine coupling constant $\alpha$) which are independent of the contour:

$$Z_2 = e^{-a \ln(\mu b)} \quad \text{and} \quad \gamma_F(\alpha, \lambda) = \frac{\mu}{Z_2} \frac{\partial Z_2}{\partial \mu} = -a = -\frac{\alpha}{2\pi}(3 - \lambda). \quad (9)$$

Accordingly, the full, renormalized one-fermion Green function is [6,12]

$$G_{ren}(x,y) = \mu^a \int \frac{d^4p}{(2\pi)^4} e^{-ip\cdot(x-y)} \frac{\Gamma(1+a)}{(i\not{p}+m_r)^{1+a}}, \qquad (10)$$

which makes apparent the infraparticle structure of the physical fermion [13].

## 4. Universal Infraparticle Vertex Function

In the light of the preceeding discussions it seems useful to consider a nontrivial contour with an obstruction, notably a self-intersecting loop at some point $z$. This contour corresponds to the following generic process: an infraparticle on route along a smooth contour with four-velocity $u_1(\tau)$ is suddenly derailed at $z$ and then proceeds to propagate again on another smooth contour characterized by four-velocity $u_2(\tau)$. For the sake of simplicity and with no loss of generality, both smooth contours can be identified with straight lines (which are the RG asymptotes). Such a process can be adequately described by an appropriate three-point function $\Gamma(x,y,z)$ which accounts for the four-velocity change at $z$, and which labels initial and final states by four-velocities. The purpose of the infinitesimal self-intersecting loop is to *simulate* the overlap of the soft boson "clouds" accompanying the charged fermion during the four-velocity transition process. This is a universal process and, under certain circumstances, it may not depend on the particular group character of the soft interactions. Being an effective geometric approach, relying on contours, this treatment can only provide a *global* formulation of soft interactions, while specific *local* features are left out, limiting the validity range of the method. It is, however, realistic to follow this assessment and calculate a vertex function which enables the emulation of the Isgur-Wise form factor [14] in the physically accessible kinematic regime [1,2,15].

To this end, we define the three-point function

$$\Gamma(x,y,z) \stackrel{c\to 0}{=} \int_c^\infty dT \int_0^T ds \int [dp(\tau)] \, \text{tr}\mathcal{P} \exp\left\{\int_0^T d\tau \, [ip(\tau)\cdot\gamma + m]\right\} G(x,y,z), \qquad (11)$$

where

$$G(x,y,z) = \int_{\substack{x(0)=x\\x(s)=z\\x(T)=y}} [dx(\tau)] \exp\left[i\int_0^T d\tau\, p(\tau)\cdot\dot{x}(\tau)\right] \left\langle \exp\left\{ig\int_0^T d\tau \dot{x}(\tau)\cdot A[x(\tau)]\right\}\right\rangle_A. \qquad (12)$$

The next step is to split the interaction point according to $x(s) \to [x(s_1), x(s_2)]$, hence generating an entrance and an exit point at the vertex $z$, at mutual distance of the order of the ribbon width $b$, and compensating for the doubling induced on the integration measure, i.e., $d^4x(s) \to d^4x(s_1)d^4x(s_2)$ via $\delta[x(s_1) - x(s_2)]$. The net effect of the integration constraints, implemented through the product of delta functions $\delta[x(0)-x]\,\delta[x(T)-y]\,\delta[x(s_1)-x(s_2)]\,\delta[x(s_1)-z]$, is to rearrange the contour involved in the evaluation of the three-point function into three disjoint branches, one of which forms a self-intersecting loop $C_{zz}$ at $z$.

Under these contour specifications, the boson correlator acquires the form
$\left\langle \exp\left[ig \int_{L_{x,y}^z} dw_\mu \, A_\mu(w)\right]\right\rangle_A \left\langle \exp\left\{ig \int_{s_1}^{s_2} d\tau \, \dot{x}(\tau) \cdot A[x(\tau)]\right\}\right\rangle_A$. This is done by attaching the disjoint contour segments $L_{zx}$ and $L_{zy}$, corresponding to four-velocities $u_1$ and $u_2$, at the interaction point $z \pm 0^+$, where the four-velocity becomes discontinuous. An appropriate coordinate rendering of $L_{x,y}^z$ in the interval $[0,1]$, with the interaction point $z$ removed, is

$$x^L(t) = \begin{cases} x\left[t(T - \tilde{\tau})\right], & 0 \leq t < \tilde{t} \\ x\left[t(T - \tilde{\tau}) + \tilde{\tau}\right] & \tilde{t} < t \leq T, \end{cases} \quad (13)$$

where $\tilde{t} = s_1/(T - \tilde{\tau}) = (s_2 - \tilde{\tau})/(T - \tilde{\tau})$ with $\tilde{\tau} = s_2 - s_1$ and $x^L(\tilde{t}) = z$, the interim point $\tilde{t}$ beeing excised from the interval $[0, 1]$.

Because the boson correlator factorizes, the three-point function becomes the product of a purely kinematical factor $G(x, y, z)$ which describes coexisting in and out four-velocity states (see for details [2]), and a factor $G^{(i)}(z, z)$ which encodes information how the dynamical "pinch" at $z$ derails the infraparticle and causes a discontinuity in the four-velocity, i.e., $G^{(i)}(x, y, z) = G(x, y|z) \, G^{(i)}(z, z)$. The zeroth order term ($i = 0$) constitutes an overall kinematical factor, $G^{(0)}(z, z) = 1$, associated to a loop that shrinks to zero upon the identification $s_1 = s_2$. The first nontrivial contribution comes from the factor $G^{(1)}(z, z)$ which corresponds to a self-crossing loop (see Fig. 1) in the particle's contour to be entered and exited *smoothly* with four-velocities $u_1$ and $u_2$, respectively. It thus appears that this term is responsible, in a geometric sense, for the transport of an intact soft boson "cloud" through the interaction point.

A uniform parametrization of the loop integral $\oint_{C_{zz}^{(1)}} dx_\mu A_\mu(x)$ is provided by $\tau \to t = \tau - (s_1 + s_2)/2$ accompanied by the coordinate readjustment $x(\tau) \to x^c(t) = x\left[+ (s_1 + s_2)/2\right]$, where $t$ is restricted in the interval $[-\frac{\tilde{\tau}}{2}, \frac{\tilde{\tau}}{2}]$ and $\tilde{\tau} = s_2 - s_1$. Then, all gauge-dependent terms vanish identically, as expected, and employing again the ribbon regularization, the final result for the boson line exponential is [2]

$$I(\theta) \simeq - \left(g^2/4\pi^2\right) \ln\left(\tilde{\tau}/2b\right) \pi\theta/\sqrt{1 - \theta^2}, \quad (14)$$

where $\theta = u_1 \cdot u_2$. In order to disentangle IR from UV effects, we make a RG scale readjustment to tune the coupling constant to values relevant in the IR domain: $g^2(b^2) \ln(\tilde{\tau}/2b) = g^2(\mu^2) \ln(\tilde{\tau}\mu) = g^2(\mu^2) \ln k$. One may view $k$ as the parameter which relates the scales characterizing long- and short-distance regimes. The result for the three-point function after renormalization is

$$\Gamma_{ren}^{(1)}(x, y, z) = -\exp\left(-\frac{g^2}{4\pi} \frac{\theta}{\sqrt{1 - \theta^2}} \ln k\right) \Gamma_{ren}^{(0)}(x, y, z). \quad (15)$$

The nonperturbative part of the vertex function is obtained by removing kinematical contributions according to

$$F(x, y, z) = \frac{\Gamma_{ren}^{(0)}(x, y, z) + \Gamma_{ren}^{(1)}(x, y, z)}{\Gamma_{ren}^{(0)}(x, y, z)} \quad (16)$$

to get the following *universal* (i.e., mass- and contour-independent) expression

$$F(x,y,z) = F(\theta) = 1 - \exp\left(-\frac{g^2}{4\pi} \frac{\theta}{\sqrt{1-\theta^2}} \ln k\right), \qquad (17)$$

which can be analytically continued to Minkowski space by replacing $\theta \to w \equiv \tilde{u}^1_\mu \tilde{u}^{2\mu} = \frac{1}{\theta}$, where $\tilde{u}^1$ and $\tilde{u}^2$ are the Minkowski counterparts of $u^1$ and $u^2$, respectively [2]. The connection to the leading-order Isgur-Wise form factor is established by replacing $e^2$ by $(4/3)g_s^2$ and adapting the parameter $k$ to scales involved in heavy-meson decays by setting $k = \frac{m_Q}{\bar{\Lambda}}$, where $m_Q$ is the heavy-quark mass, and $\bar{\Lambda}$ characterizes the energy scale of the light degrees of freedom. Then the final result reads

$$\xi(w) = 1 - \exp\left(-\frac{4}{3}\alpha_s \frac{1}{\sqrt{w^2-1}} \ln k\right). \qquad (18)$$

In the kinematic region accessible to semileptonic decays ($1 \leq w \leq 1.6$), and using rather typical values of the involved parameters [4], *viz.*, $k \approx 8$ and $\alpha_s \simeq 0.21$, this first-principles vertex function yields remarkable agreement with recent data of different experimental groups (for an explicit comparison, see [1,2]).

## 5. Conclusions

This article tries to draw a unifying view on *universal* features of soft interactions using an approach which lends itself by construction to the renormalization group and relies upon particle path integrals. Arguments are given how to pave the way for generalizations of the framework and extensions to the UV regime.